# Water rotational jump driven large amplitude molecular motions of nitrate ions in aqueous potassium nitrate solution

Puja Banerjee, S. Yashonath[*], and B. Bagchi[*]

Solid State and Structural Chemistry Unit,
Indian Institute of Science, Bangalore 560012, India

# Abstract

Molecular dynamics simulations of aqueous $KNO_3$ solution reveal a highly complex rotational dynamics of nitrate ions where, superimposed on the expected continuous Brownian motion, are large amplitude angular jumps that are coupled to and at least partly driven by similar large amplitude jump motions in water molecules which are associated with change in the hydrogen bonded water molecule. These jumps contribute significantly to rotational and translational motions of these ions. We explore the detailed mechanism of these correlated (or, coupled) jumps and introduce a new time correlation function to decompose the coupled orientational-jump dynamics of solvent and solute in the aqueous electrolytic solution. Time correlation function provides for the unequivocal determination of the time constant involved in orientational dynamics originating from making and breaking of hydrogen bonds. We discover two distinct mechanisms-both are coupled to density fluctuation but are of different types.

## 1. Introduction

The solvation and solvent dynamics have been investigated in considerable detail in the recent past[1]. Hydrogen bond dynamics is important in many biological and chemical processes. Since the discovery of jump orientational dynamics of water instead of a continuous water reorientation by Laage-Hynes[2,3], there has been an increased interest in such behaviour. Laage-Hynes also studied reorientational dynamics of water around hydrophilic solutes (such as monoatomic anions $Cl^-$, $Br^-$, $I^-$ and amino acids [4-6]) and some hydrophobic solutes [7] using molecular dynamics simulation. Many experimental studies investigating water dynamics have been reported[8-10].

Orientational and translational motion of oxyanions like $NO_3^-$ plays an important role in such dynamics. In an earlier molecular dynamics study, Barrat and Klein studied the fused salt $[Ca(NO_3)_2]_{0.4}[KNO_3]_{0.6}$, and found the jump dynamics of nitrate ion similar to those seen in the plastic crystalline phase[11,12]. Whittle and Clarke studied calcium nitrate solution and found that, at higher concentrations, there is a broadening of the depolarized Rayleigh scattering and attributed the reorientation of nitrate ion to associated ion pairs [13]. Thogersen et al investigated the hydration shell around the nitrate ion with the help of 2D-IR, UV-IR and UV-UV time-resolved spectroscopies. They found that the hydration shell is labile and the $D_{3h}$ symmetry of the ion is broken in water and the degeneracy is lifted [14].

*yashonath@sscu.iisc.ernet.in
*bbagchi@sscu.iisc.ernet.in

Recent study by Xie et al. reports two reorientatinal relaxation routes of water molecule in nitrate hydration shell 15. The authors have observed water switching hydrogen bonds between two oxygens of the same nitrate as well as between a nitrate oxygen and another water oxygen. They further discuss the effect of dual reorientation relaxation routes on the residence time as well as the rotation of nitrate in water.

 Here we investigate the solvation as well as solute dynamics of $KNO_3$ solution by molecular dynamics(MD) simulation. The structure of the hydration shell is reported and compared with both experimental and previous theoretical studies. Hydrogen bond exchange of a hydrated water molecule around the nitrate ion is seen to occur via two distinct mechanisms (i) replacing the present hydrogen bond with O of $NO_3$ by another O of same $NO_3$ or (ii) by water oxygen, similar to that reported by Xie et al. In the latter case, the earlier water hydrogen bonding to nitrate leaves the hydration shell altogether. Further we find a large fraction of hydrogen bond exchanges are only transient leading to no real H-bond exchange. Importantly, hydrogen bond exchanges are accompanied by large rotational jumps of water. Interestingly we also observe large rotational jumps of nitrate ions. The nature of these rotational jumps of nitrate ion are different for the two cases of hydrogen bond exchange of hydrated water mentioned above. Finally, it is seen that an exchange of the hydrogen bonded water interacting with the oxygen of the nitrate ion is coupled with large amplitude angular jump motion of both nitrate ion and the hydrogen bonded water molecule. We define and obtain time correlation function to understand the coupling.

## 2. Methods

$NO_3$ is taken as triangular planar shape with N-O distance of 1.226 Å. Rigid nonpolarizable force field parameters have been used for water as well as nitrate ion. SPC/E model[17] has been employed for water. For potassium ion, the OPLS-AA[18] force field and for nitrate ion, the potential model suggested by Vchirawongkwin et al have been employed[16]. The self interaction parameters are listed in Table 1 and consist of Lennard-Jones and Coulombic terms. Molecular dynamics (MD) simulations have been performed with the DL_POLY[19] package on a system of four $K^+$ and four $NO_3^-$ in 2040 water molecules in a cubic simulation cell of length 39.43 Å. This corresponds to 0.109 M concentration of aqueous $KNO_3$ solution and a density of 1.00555 g/cm3. Simulations were carried out in the microcanonical ensemble with periodic boundary conditions with a cut-off radius of 18 Å. The long-range forces were computed with Ewald summation[20,21]. Trajectory was propagated using a velocity Verlet integrator with a time step of 1 fs. The aqueous $KNO_3$ system was equilibrated for 300 ps at 300 K in the canonical (NVT) ensemble and then a 1 ns MD trajectory was generated in the microcannonical (NVE) ensemble. The coordinates, velocities and forces were stored every 5 fs for subsequent use for the evaluation of various properties.

**Table 1** Force field parameters

| Species | Atom, i | $\sigma_{ii}$ (Å) | $\varepsilon_{ii}$(Å)(kj/mol) | $q_i$ (e) | ref |
|---------|---------|---------|---------|---------|-----|
| Water | $H^w$ | 0.000000 | 0.000000 | +0.4238 | 17 |
| Water | $O^w$ | 3.166000 | 0.650000 | −0.8476 | 17 |
| Cation | K | 4.934630 | 0.001372 | 1.000 | 18 |
| Nitrate Anion | N | 3.150000 | 0.711300 | 1.118 | 16 |
| Nitrate Anion | O | 2.850000 | 0.836800 | -0.706 | 16 |

## 3 Result and discussion

### 3.1 Structure

Figure 1 shows the radial distribution functions between N and O atoms of the nitrate species on the one hand and O and H atoms of the water on the other (indicated by superscript w) for the solution at 300 K. From the N-$O^w$ rdf, it is clear that the first hydration sphere of water around the nitrate ion extends upto 4.15 Å.

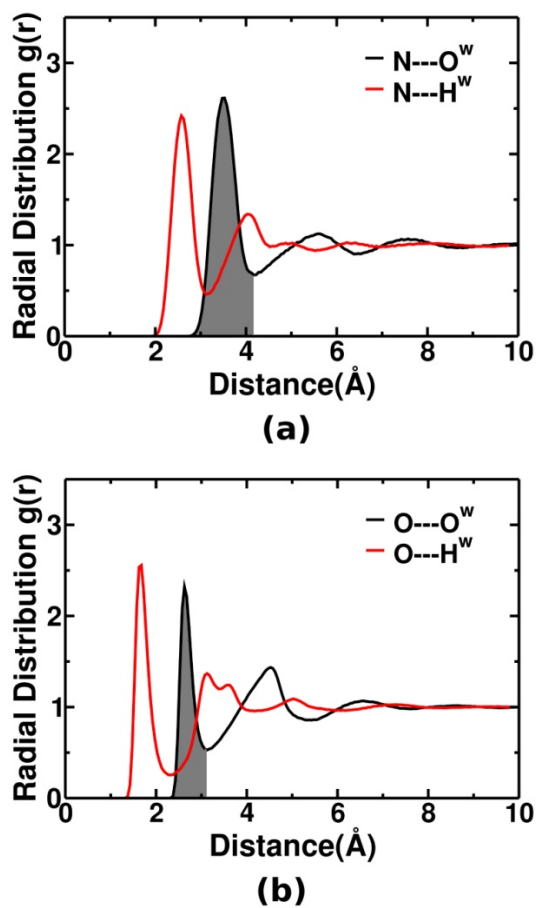

**Fig. 1** (a) Distance from nitrate nitrogen (N) to water oxygen ($O^w$) and hydrogen ($H^w$); Shaded area shows number of water oxygens around nitrate nitrogen(~ 9.2), (b) Distance from nitrate oxygen(O) to water oxygen($O^w$) and hydrogen($H^w$). Shaded area shows number of water oxygens around nitrate oxygen atom (~2.5).

In Figure 1(b) the peak between oxygen of nitrate (O) and hydrogen of water ($H^w$) seen around 1.6 Å corresponds to the hydrogen bonded pair. The second peak is split, usually a characteristic of glasses, but here it arises from the presence of non hydrogen bonded pairs of O and $H^w$ atoms in the hydrogen bonded pairs of water and nitrate ion(shown in Figure 2). Here both the other oxygens of the nitrate and the hydrogens of water are responsible for the split peak and has its origin in the molecular geometry of the nitrate and water molecules.

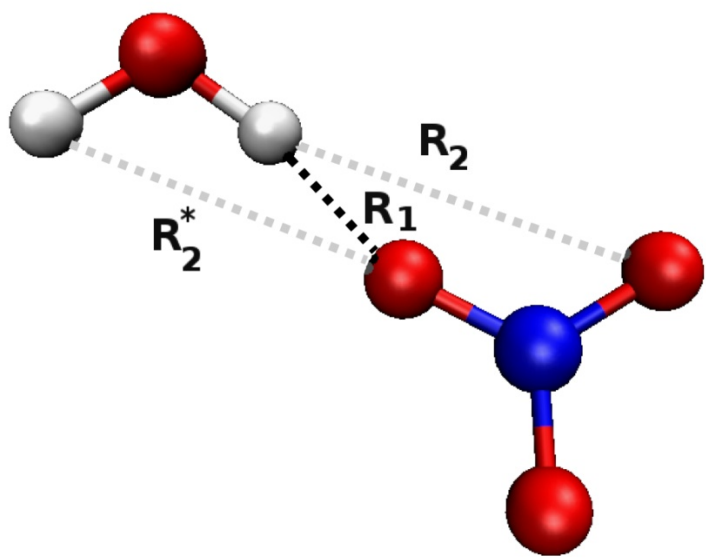

**Fig. 2** Distance R1 corresponds to the 1st peak in Figure 1b (O–$H^w$ curve) for the hydrogen bonded pairs and distance $R_2$ and $R_2^*$ corresponds to the split second peak in that curve for two non-hydrogen bonded oxygen-hydrogen pair.

The first maximum in the N-$O^w$ rdf ($R_{N-O^w}^{max}$) is at 3.5 Å. This may be compared with 3.4 Å from the work of Dang et al.[23], 3.65 Å by Xie et al.[15] and 3.5±0.311 Å obtained from X-ray scattering experiment of Pinna and coworkers[22]. The first maximum in N-Hw rdf ($R_{N-H^w}^{max}$) is at 2.6 Å. This may be compared with 2.4 Å and 2.75 Å from Dang et al. and Xie et al. respectively.

These results are summarized in Table 2 and suggest that intermolecular interaction potential employed by us in this study provides results in good agreement with earlier experimental as well as simulation results. The number of water oxygens around the nitrate nitrogen obtained by integration of the rdf is 9.2 (Figure 1(a)). This may be compared with coordination number of 10±1 from Dang et al. and 12 from Xie et al (Table 2). The number of oxygen atoms around a specific nitrate oxygen is around 2.5 oxygen atoms (Figure 1(b)). On the average, two or three water molecules can be hydrogen bonded to each of the three nitrate oxygens. Figure 3 shows the angular radial distribution(ARD) function plot of water molecules surrounding a nitrate ion. This ARD analysis clearly establish a picture of symmetric solvation shell around nitrate ion.

**Table 2** Comparison of experimental and calculated structural properties of aqueous nitrate solution. $R_{N-O^w}^{max}$ and $R_{N-H^w}^{max}$ are the positions of first maxima on the radial distribution function plots of nitrate nitrogen with water oxygen (N-O$^w$ rdf) and nitrate nitrogen with water hydrogen (N-H$^w$ rdf) respectively. Coordination number is the number of water oxygens around a nitrate nitrogen.

|  | Experimental[22] | Our Work | Dang et al.[23] | Xie et al.[15] |
|---|---|---|---|---|
| $R_{N-O^w}^{max}$, Å | 3.5±0.311 | 3.5 | 3.4 | 3.65 |
| $R_{N-H^w}^{max}$, Å | – | 2.6 | 2.4 | 2.75 |
| Coordination no. | – | 9 | 10±1 | 12 |

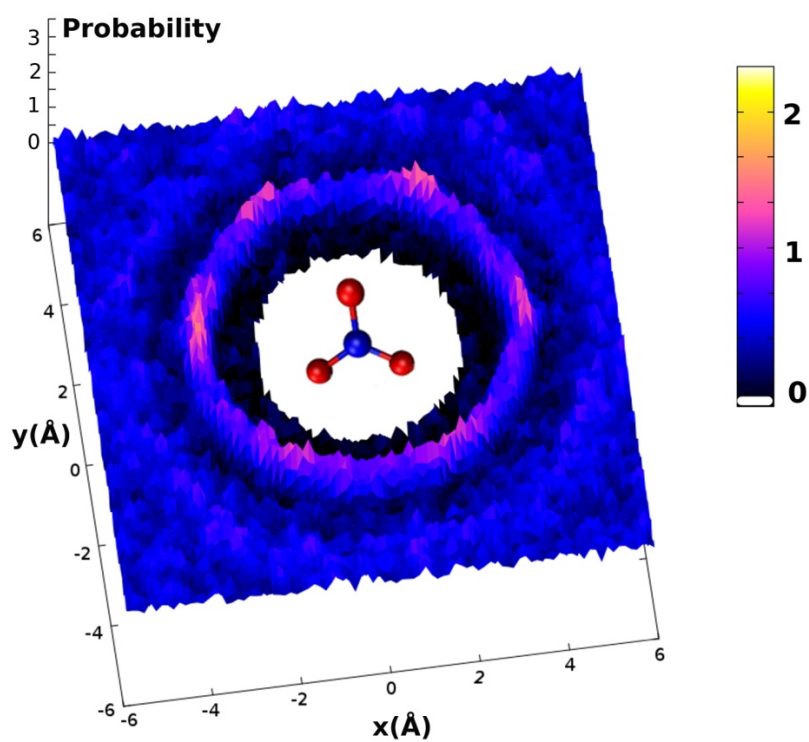

**Fig. 3** The Angular Radial Distribution function of water surrounding a nitrate ion. A color mapped on $g_{N-O^w}$ (r) shows water oxygen atoms(O$^w$) in the first solvation shell around three nitrate oxygen atoms. The high probability regions are shown in pink. Note that these are located symmetrically around the oxygen on either side of the radial vector.

In order to define a H-bond (hydrogen bond), we computed the rdf and angular distributions (Figure 4). From the plots it is evident that an appropriate definition for the hydrogen bond O$^w$H$^w$-O(hydrogen bond between nitrate ion and water molecule) is: R$_{OHw}$< 2.35 Å, R$_{OOw}$< 3.15 Å, θ$_{HwOOw}$< 24$^0$, where R$_{OHw}$ is the distance between nitrate oxygen atom (O) and water hydrogen atom ($H^w$), $R_{OO^w}$ is the distance between nitrate oxygen atom (O) and water oxygen atom ($O^w$) and $\theta_{H^wOO^w}$ is the angle between the $OH^w$ and $OO^w$ vectors.

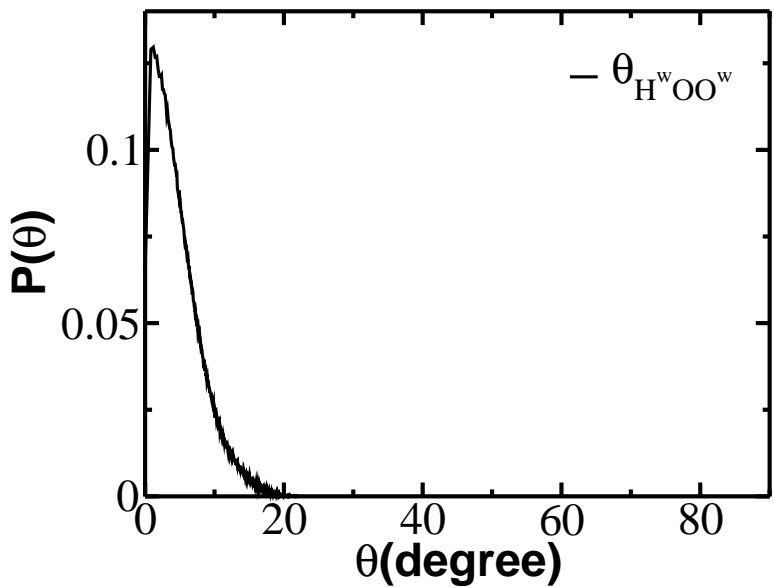

**Fig. 4** Angular distribution of the angle $\theta_{H^wOO^w}$ in the 1st solvation shell of nitrate ion. ($H^w$ :water hydrogen; O :nitrate oxygen; $O^w$ : water oxygen)

### 3.2 Dynamics of Hydrogen bond switching
### 3.2.1 Diffusion
We compute diffusion constant, D of $K^+$ and $NO_3^-$ in water from our simulations through the ion mean-square displacement by Einstein's diffusion equation:

$$D = \lim_{t \to \infty} 1 / (6t) \left\langle \left| r(t) - r(0) \right|^2 \right\rangle \qquad (1)$$

A comparison of our result (0.109 M aqueous $KNO_3$ at 300 K) with that of experiments (ions in water at infinite dilution at 298 K) , shown in Table 3 suggests that there is an good agreement between our result and experimental measurements.

**Table 3** Self diffusion constants of ions in water.

|  | Experimental[24] | Our work |
|---|---|---|
| $D_{K+}$, ($10^{-5}$ cm$^2$/s) | 1.957 | 1.22 |
| $D_{NO3^-}$, ($10^{-5}$ cm$^2$/s) | 1.902 | 1.24 |

### 3.2.2 Reorientational motion of solute and solvent

When hydrogen bond between nitrate ion and water is broken and the water hydrogen hydrogen bonded (H-bonded) to nitrate oxygen until now hydrogen bonds to another atom then we refer to it as a Hydrogen Bond Switching event (HBS event) of water surrounding nitrate ion. Previously, Laage-Hynes suggested that the reorientation of water during hydrogen bond switching events (HBS event) occurs via large amplitude angular jump[2]. Recently, Xie et al. studied jump dynamics of water in the first hydration shell of nitrate ion[15].

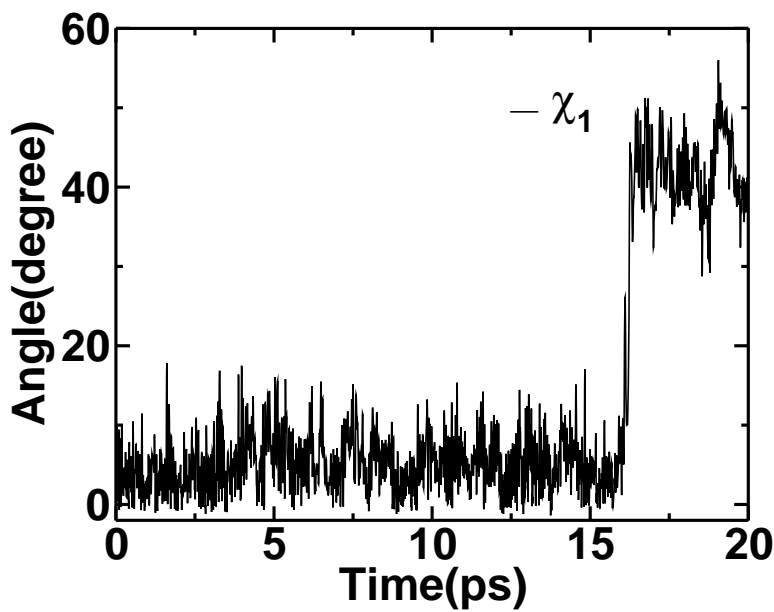

**Fig. 5** Time evolution of $\chi_1$ (the angle between a selected O-H bond which was initially H-bonded to nitrate and laboratory frame axis) during H-bond switching event.

Water in the hydration shell around the nitrate shows both librational motion as well as diffusive motion. We focus on the reorientation of water molecules in the first hydration shell of nitrate ion. A cursory observation showed that the angle of the hydrogen bonded O-H group makes with the laboratory frame z-axis changes abruptly during a HBS event. In Figure 5, we show the variation of the angle that the O-H group makes with laboratory frame axis ($\chi_1$) as a function of time. Here time 0 is the time when that chosen O-H had a stable hyderogen bonding with a NO group. It is seen that angle between O-H bond and reference laboratory frame changes abruptly at 15 ps. This large change in the angle suggests jump reorientational motion of water during hydrogen bond breaking.

The nitrate ion during H-bond exchange also shows interesting behavior. Careful observation shows that nitrate N-O bond also shows a jump rotational dynamics during HBS event. Figure 6 shows a jump in the angle between the N-O group and laboratory frame axis ($\chi_2$) during hydrogen bond switching.

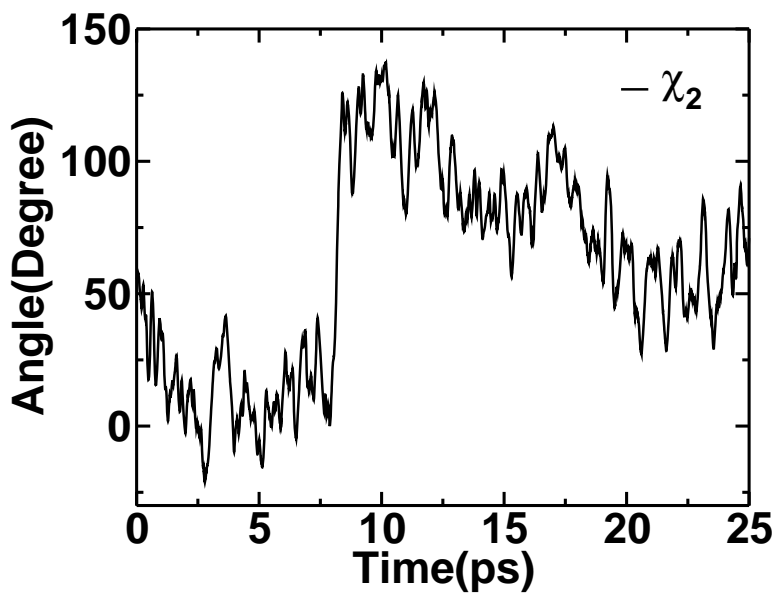

**Fig. 6** Time evolution of $\chi_2$ (the angle between a selected N-O bond and laboratory frame axis) during H-bond switching event, that is, when the water to which this particular O of $NO_3^-$ ion was H-bonded changed.

An orientation analogue of van Hove self-correlation function ($G(\psi,t)$) in which distance r is replaced by angle $\psi$, is plotted for nitrate ions surrounded by water molecules (Figure7). $G(\psi, t)$ is defined as

$$G(\psi,t) = \left\langle \delta\left[\psi_i - \psi_i(t)\right] \right\rangle \qquad (2)$$

Where
$$\psi_i(t) = \cos^{-1}\left[\vec{u}_i(0) \cdot \vec{u}_i(t)\right] \qquad (3)$$

and $\vec{u}_i(t)$ is the N-O bond vector at time t. Figure 7 is obtained by averaging over all time and all nitrate ions.

It is seen that even after about just 2 ps delay, a small peak begins to appear around $110^0$. By 40 ps there is significant intensity around $110^0$. By 80 ps, the intensity of this peak has reached a maximum while the intensity of the peak between 0-30 degrees has decreased significantly. These two well separated peaks at longer times indicate a large amplitude angular jump of less mobile nitrate ion[25]. Note that the diffusivity of the nitrate ion is lower than that of water : $D_{NO_3^-}$ = $1.24*10^{-5}$ cm$^2$/s as compared to $D_{H2O}$ = $2.87*10^{-5}$ cm$^2$/s. The non-zero intensity of $G(\psi, t)$ in between two peaks arises from the fact that there are also simple diffusive motion of nitrate ion which are not showing the reorientational jump.

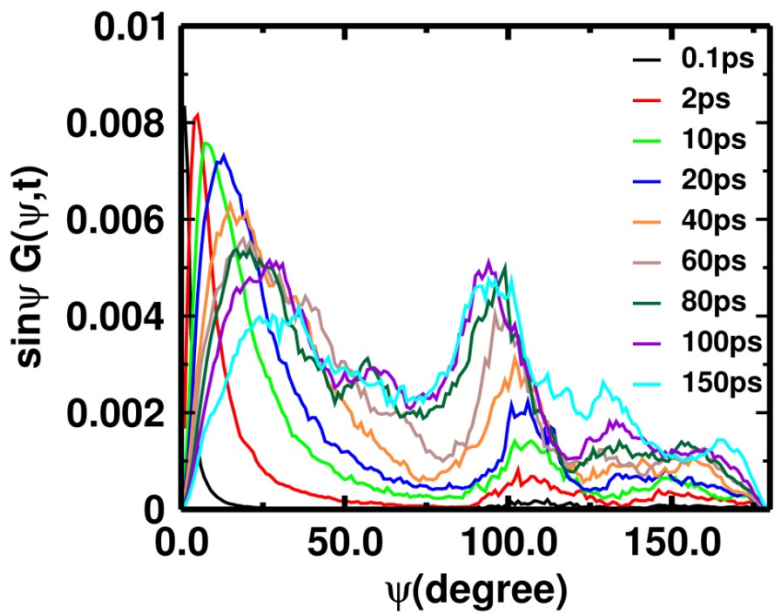

**Fig. 7** Angle distribution (sin ψ G(ψ, t)) associated with self part of the van Hove correlation function for N-O bond vector of nitrate in aqueous solution at various times. The two peak character of the distribution at longer times indicates an angular jump of approximately around $100^0$ of N-O bond in aqueous solution.

### 3.2.3 Correlation between jump motion of nitrate and water

We have seen that both nitrate ion as well as the water in the hydration shell of nitrate ion perform large amplitude jump rotations during a HBS event. We now explore the relation between the (i) hydrogen bond switching(HBS), (ii) the jump dynamics of water and (iii) the jump dynamics of nitrate ion. In Figure 8, we show the angle that water as well as nitrate makes with the laboratory z-axis. The zero time is the time when hydrogen bonding was there between that water O-H and nitrate N-O. In Figure 8(a), the HBS event is followed by simultaneous jump rotation of both water and nitrate ion. On the other hand, in Figure 8(b), you can see that the nitrate ion performs the jump rotation after ~1 ps of water jump. In general, we have observed that the water and the nitrate ion perform jump rotation within a few picoseconds of the HBS event. Thus, we can say that, the HBS event triggers the jump rotation of both the water molecule (which was H-bonded to nitrate ion but breaks its H-bond) and the nitrate ion. Figure 9 shows the sequence of events including breaking of a hydrogen bond between nitrate oxygen and water hydrogen. If you observe carefully, you can see that at first water molecule has reoriented and one H-bond has broken. This is followed by the out-of-plane reorientation of nitrate ion.

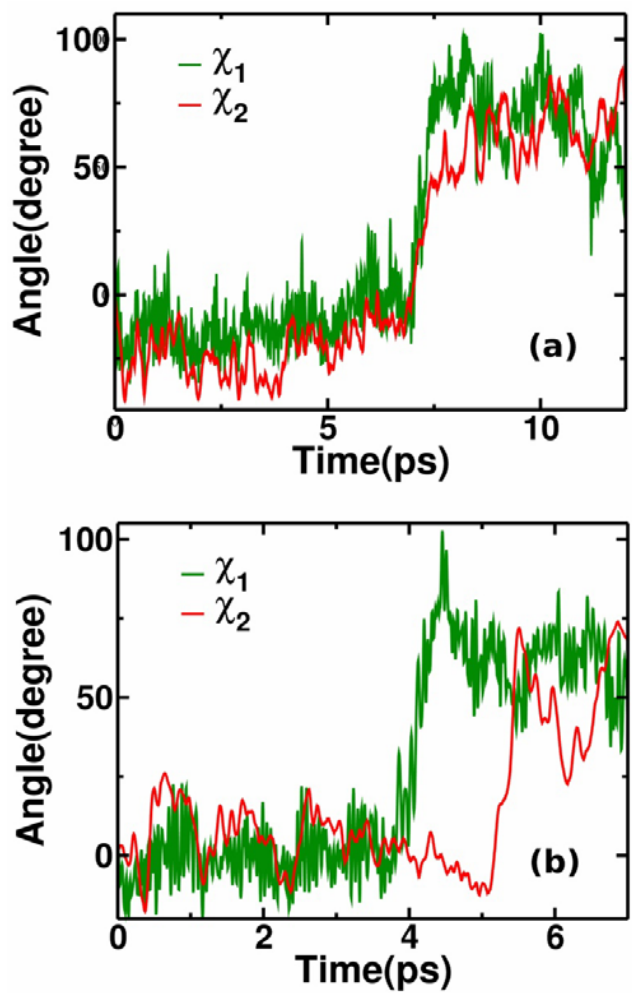

**Fig. 8** Time evolution of $\chi_1$ (the angle between a selected O-H bond and laboratory frame axis) and $\chi_2$ (the angle between a selected N-O bond and laboratory frame axis) during H-bond switching event of two selected O-H and N-O bonds. Note that in (a) both bonds reoriented simultaneously while in (b) there is a time lag between two angular jumps.

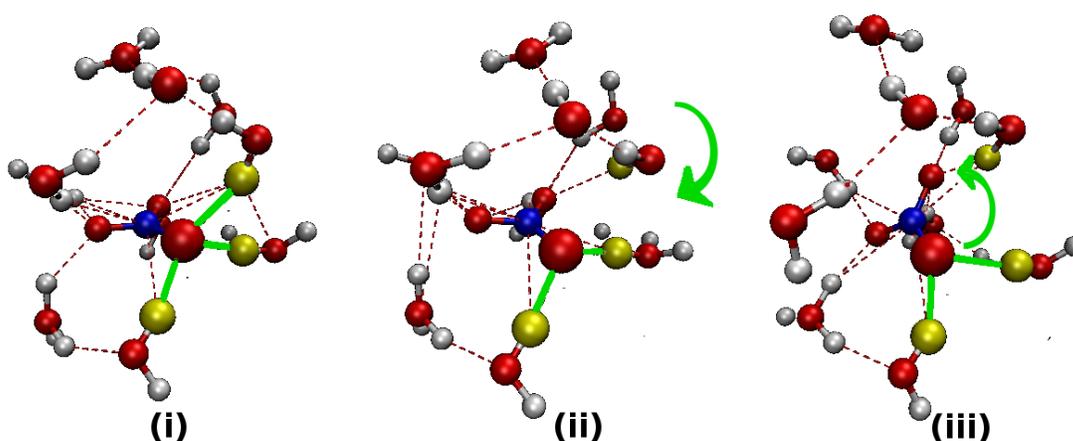

**Fig. 9** Snapshots of the first hydration shell around the nitrate ion during a hydrogen bond switching event: (yellow atoms are H-bonded hydrogens at 0 ps) (i) 0 ps (3 H atoms are H-bonded to one of the nitrate oxygens) (ii) 0.05 ps (1 H-bond broken; water O-H bond reorients) (iii) 0.65 ps (nitrate molecule has an out of plane rotation)

### 3.3 Mechanism of hydrogen bond switching

To determine the correlation between these two jump events of nitrate ion and water during a HBS event, a detailed analysis of hydrogen bond switching event between a nitrate ion and surrounding water will be of help. To analyze the angular jump of N-O and O-H following a hydrogen bond switching event, we need to look into the molecular mechanism of hydrogen bond switching. To characterize a HBS event, we designate the participant atoms as follows: A particular N*-O* bond breaks H-bond with $H_a$ and makes a new H-bond with $H_b$ where $H_a$ and $H_b$ belong to two different water molecules $a$ and $b$.

Ha of old donor water molecule makes a new H-bond with an acceptor $O_c$ after it breaks H-bond with O*. This is shown in Figure 10. We now look rotation of N*-O* and $O_a$-$H_a$. To obtain the average behavior we have collected and analyzed 9151 such H-bond switching events and also checked the results with different initial configurations. The properties reported below are averages over these large number of HBS events. We have monitored the distance $R_{N*O_a}$, the angle $\theta$ between the projection of N*-O* on the plane of $H_aN*H_b$ and the bisector of angle $H_aN*H_b$ and the angle $\varphi$ between the projection of $O_a$-$H_a$ on the plane of $O*O_aO_c$ and the bisector of angle $O*O_aO_c$. Thus, $\theta = 0^0$ and $\varphi = 0^0$ gives the time when the corresponding bonds coincide with the bisector. We define this as the time when the rotational jumps of N*-O* and $O_a$-$H_a$ take place. Here time 0 is different for these three curves: for $R_{N*O_a}$, it is the time of H-bond breaking between a particular N-O and O-H, for $\theta$, it is the time of that N-O jump of nitrate ion and for $\varphi$, it is O-H jump time of water. These three times for a HBS event may be same or may not. We found that depending on the type of $O_c$ atom, there exist different types of mechanisms.

**Mechanism 1**: This is characterised by $O_c$ which is a water oxygen, $H_a$ switches H-bond from O* to $O_c$ (35% of all HBS event analysed) (Figure 10(a)). We have observed that, subsequently, water molecule $a$ moves out of first hydration shell of the nitrate ion. This is seen from the distance $R_{N*O_a}$ which increases (Figure 10b) and becomes greater than the cutoff distance (4.15 Å see Fig. 1(a)) after HBS event. This mechanism is characterised by average magnitude of angular jump of N*-O* bond ($\theta$) of $110^0$ and that of Oa-Ha bond ($\varphi$) is around $50^0$.(Figure 10(b)).

**Mechanism 2**: Here $O_c$ is not a water oxygen but instead $O_c$ is one of the nitrate oxygen of the same nitrate ion. However, there are two possibilities within this mechanism : (i) in which the $O_c$ happens to be the same O* as the oxygen to which Ha was H-bonded. That is, a transient breaking of H-bonds can occur frequently and Ha again forms H-bond with the same O* after that transient breaking(40% of all HBS event analysed). But our interest here is to analyse the situation when (ii) $O_c$ is a nitrate oxygen of the same nitrate ion other than the previously Hbonded O of nitrate (O*).(25% of all HBS event analysed) (Figure 10(c)). Here the distance $R_{N*O_a}$ does not increase as $O_a$ remains in the first hydration shell after HBS event (Figure 10(d)). In this case, the average magnitude of angular jump of N*-O* bond ($\theta$) and $O_a$-$H_a$ bond ($\varphi$) are $80^0$ and $30^0$ respectively (Figure 10(d)). Note that the magnitude of angular jumps (both $\theta$ and $\varphi$) are less than what was obtained for mechanism 1.

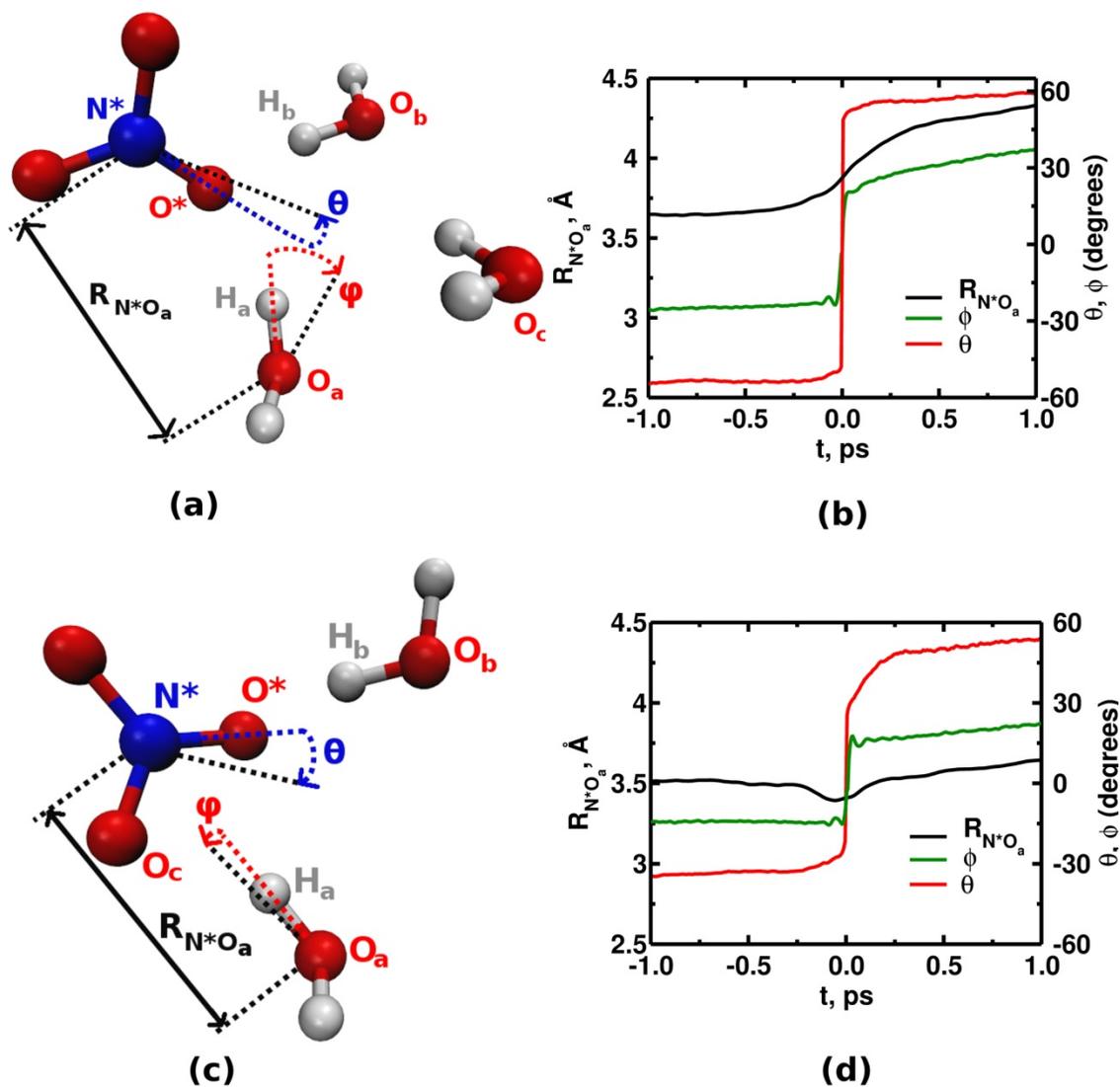

**Fig. 10** Hydrogen bond switching event of nitrate ion occurs through two distinct mechanisms. Definition of geometric coordinates used to find angular jump of O-H and N-O : distance $R_{N*O_a}$ and the angles $\theta$ (between projection of N*-O* vector on the plane $H_aN*H_b$ and bisector of angle $H_aN*H_b$) and $\varphi$ (between projection of $O_a$-$H_a$ vector on the plane $O*O_aO_c$ and bisector of angle $O*O_aO_c$). Mechanism 1: (a) Definition of geometric parameters, (b) average variation of $R_{N*O_a}$, $\theta$, $\varphi$ with time. (c) and (d) for Mechanism 2.

Xie et al. have reported both these mechanisms for water reorientation. We have here not only identified these two events for water reorientation but have also obtained the average magnitude of nitrate ion jump rotation. Later we look at the nature of nitrate ion jump. We shall see that the nature of nitrate ion jump for the two mechanisms are entirely different.

### 3.4 Time correlation function for the Coupled Jump Dynamics(CJD)

We propose a new time correlation function $C_{CJD}(t)$ which can determine the time range upto which these two angular jumps are correlated. As shown in Figure 8, these two jumps can occur simultaneously or have a time lag between them. $C_{CJD}(t)$ is defined as:

$$C_{CJD}(t) = \left\langle H\left[ \left( \delta\theta_{NO_3}(\tau) - \left( \Delta\theta_{NO_3} / 2 \right) \right) * \left( \left( \Delta\varphi_w / 2 \right) - \delta\varphi_w(\tau + t) \right) \right] \right\rangle \qquad (4)$$

where $\delta\theta_{NO3}(t)$ & $\delta\varphi_w(t)$ are the difference in the angle at time t to the value at the onset of HBS event. $\Delta\theta_{NO3}(t)$ & $\Delta\varphi_w$ are the difference in the angle at the end of the HBS event to that at the beginning of the HBS event. Angular jump of water O-H ($\Delta\varphi_w$) has a distribution shown in Figure 11(b) for both mechanisms.

In both cases, H[ ] is the Heaviside theta function defined by:

$$H[x] = \begin{cases} 0, x < 0 \\ 1, x \geq 0 \end{cases} \qquad (5)$$

Briefly, the eq.(4) is described. There are two cases: (i) in which the nitrate N*-O* rotational jump preceedes the water $O_a$-$H_a$ reorientational jump. In this case, the first factor becomes > 0 first when nitrate has reoriented and second factor is > 0 until water reorients. Thus, in this intermediate time between the rotational jump of nitrate and water H[x]=1.

In the second case, when the nitrate jump reorientation succeeds water reorientational jump, the second factor < 0 after water jump. The first factor remains < 0 until nitrate reorients. Thus, H[x]=1 in this intermediate time period.

In summary, the function H[ ] on the right hand side is zero before the rotational jump of either nitrate or water is half way and remains unity until the other molecule (nitrate or water) rotates half way. When both have rotated more than half way, then H[ ] = 0. In other words, correlation remains until both the solute and solvent have completed more than half of their respective rotations during HBS event.

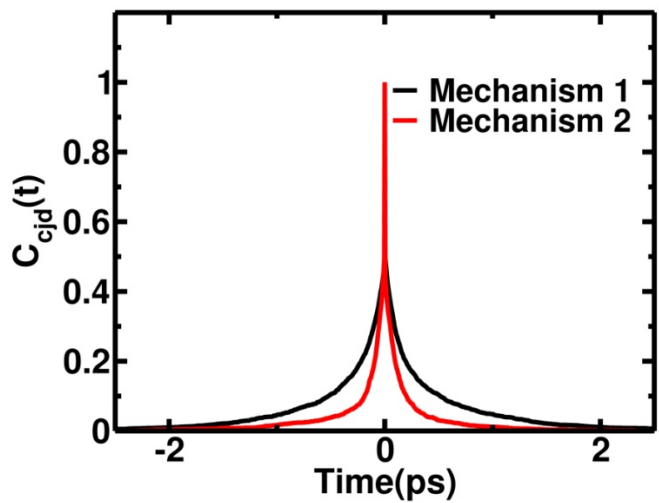

**(a)**

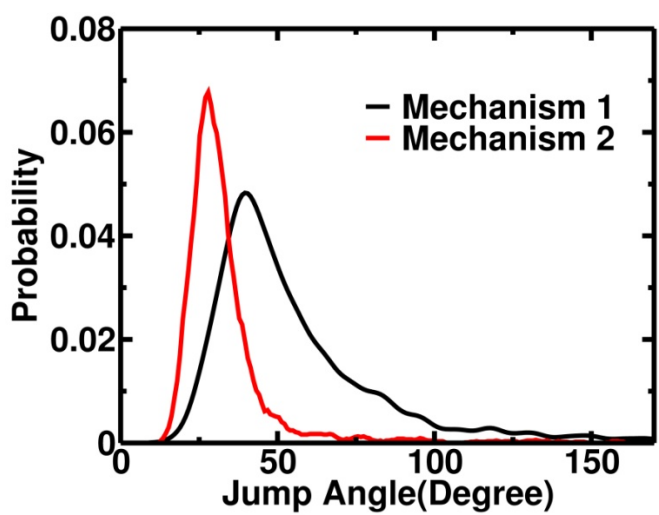

**(b)**

**Fig. 11** (a) Time correlation function ($C_{CJD}(t)$) which shows the time upto which two angular jump can be correlated; for mechanism 1 (black curve) and for mechanism 2 (red curve). (b) Distribution of Jump angle of water for mechanism 1 (black curve) and mechanism 2 (red curve)

### 3.5 Difference in the nature of reorientation of nitrate ion in the different mechanism of HBS event

Having obtained two different types of mechanisms of hydrogen bond switching in the system of aqueous nitrate, another point of interest is the nature of reorientation motion of nitrate ion for the two mechanisms of HBS event. Being a planar molecule, nitrate ion can show both in-plane reorientation and out-of-plane reorientation. To analyse the nature of rotational motion of nitrate ion exhibits in these two mechanisms, we compute the torque experienced by the nitrate ion during HBS event. We then resolve the total torque vector into two components: (a) along the $C_3$ axis and (b) along one of the $C_2$ axis.The component along $C_3$ axis provides us the torque responsible for 'cog-wheel' in-plane rotation while that along $C_2$ axis gives us the torque responsible for the rotation of the nitrate plane (out-of-plane rotation). Figure 12 shows the variation of torque with time calculated using coordinate and force of all three nitrate oxygen atoms of a nitrate ion from the trajectory file using the relation:

$$\vec{\tau} = \sum_{i=1}^{3} \vec{r}_i \times \vec{f}_i \tag{6}$$

where $\vec{r}_i$ are the vectors from N to O and $\vec{f}_i$ is the force acting on three nitrate oxygen atoms.

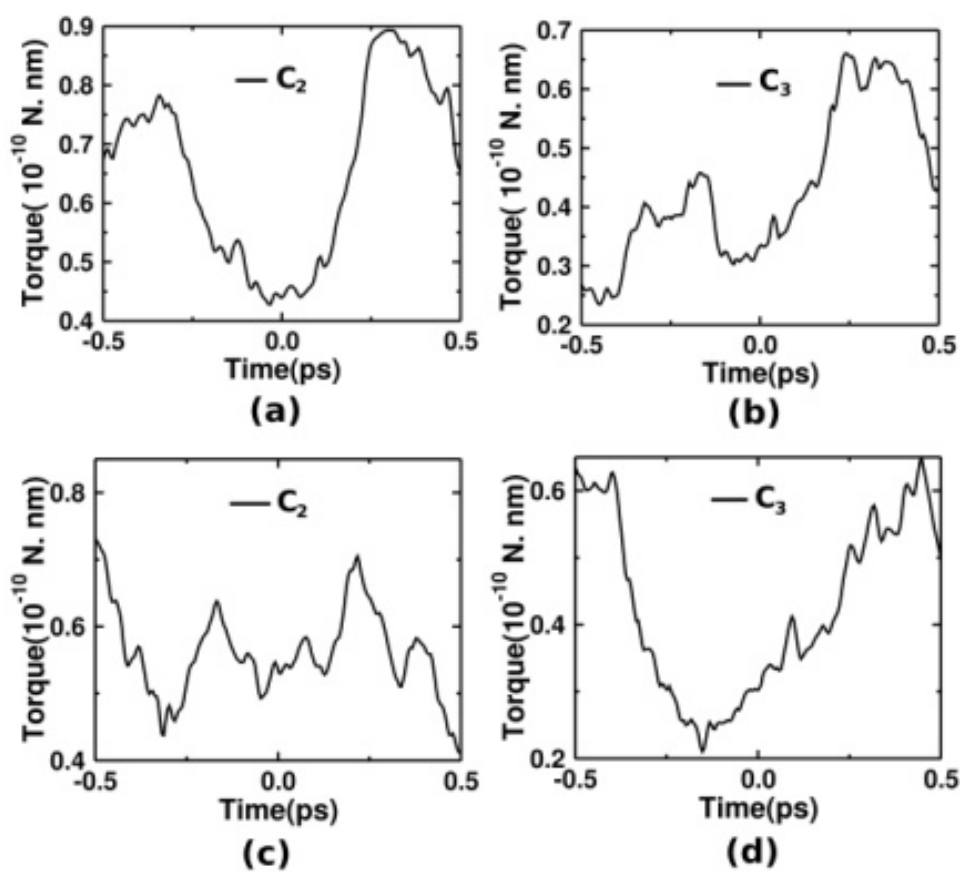

**Fig. 12** Variation of torque (acting on the nitrate nitrogen atom due to the rotational jump of nitrate during hydrogen bond switching) with time. Mechanism 1: (a) torque along $C_2$ axis of nitrate ion during HBS event, (b) torque along $C_3$ axis of nitrate ion during HBS event, (c) and (d) for Mechanism 2. Larger increase in torque along $C_2$ axis for Mechanism 1 suggests a predominance of out-of-plane rotation, while larger change of torque along $C_3$ axis for Mechanism 2 suggests in-plane rotation of nitrate ion is favorable in case of Mechanism 2 where water hydrogen exchanges H-bond between two nitrate oxygen atoms of the same nitrate ion.

Figure 12(a) and Figure 12(b) shows torque acting on nitrogen of nitrate along $C_2$ axis and $C_3$ axis respectively for a particular HBS event following Mechanism 1. The larger increase in torque acting along $C_2$ axis suggests that out-of-plane rotation of nitrate ion is predominant in Mechanism 1 where water hydrogen switches H-bond from nitrate oxygen atom to another water oxygen atom. On the other hand, Figure 12(c) and Figure 12(d) demonstrates the changes in the torque components for Mechanism 2. Here larger increase in torque along $C_3$ axis shows predominance of in-plane rotation of nitrate ion in Mechanism 2 of HBS event where water hydrogen switches H-bond from one nitrate oxygen to another nitrate oxygen of the same nitrate ion in a 'cog-wheel' kind of rotational motion. These results have been tested for other HBS events of Mechanism 1 and Mechanism 2 also.

## 4   Conclusions

In summary, it is seen that the exchange in hydrogen bonded water around a nitrate ion is accompanied by large amplitude rotational jumps of the hydrated water as well as the nitrate group. We find two distinct mechanisms of hydrogen bond exchange: (i) where the hydrated water breaks the hydrogen bond from the $NO_3^-$ and bonds with another water and (ii) where the hydrated water bonds to another oxygen of the same nitrate ion. These results are in excellent agreement with Xie et al. We, however, find that the rotational jump of water is accompanied by jump rotation of the nitrate ion, as well. Unlike in pure water, the rotation of the solute and the solvent are mutually coupled with each other and these two are, in turn, coupled to HBS event.

The origin of the coupled jump dynamics essentially lies in two factors : (i) the nitrate ion with its triangular planar arrangement, and (ii) the open framework structure of surrounding water molecules. It is unquestionably the network structure of water with locally tetrahedral arrangement that allows and sustains frequent large amplitude jumps of water because the local energy minima are present in finite spatial and orientational separations between the interacting molecules. The nitrate ion, stabilized by the hydrogen bonds to water molecules, becomes a part of the fluctuating hydrogen bond network of water and participates in the large amplitude jump motions, in addition to the usual Brownian motions.

A theoretical treatment of this coupled motion would require a calculation of the torque experienced by the nitrate ion due to the water molecules at specific relative orientations. This coupling essentially requires a mode coupling theory formulation where the torque is expressed in terms of orientation and position dependent water density relative to the central nitrate ion. Such a formulation shall naturally include the effects of jump motions of the water molecules The coupled jump motion causes faster decay of force-force and torque-torque time correlation function which in turn lower both translational and rotational friction. This can enhance the conductance.

## Acknowledgement

The work was supported by Department of Science and Technology( DST), GOVT. of India and partly by Council of Scientific and Industrial Research (CSIR), India.